\documentclass[12pt]{article}
\usepackage[dvips]{color}
\usepackage{epsfig}
\usepackage{amsmath}
\usepackage{graphicx}

\begin{document}
\title{\bf Thermodynamics of a charged hairy black hole in (2+1) dimensions}
\author{{J. Sadeghi\thanks{Email: pouriya@ipm.ir}\hspace{1mm} and H. Farahani\thanks{Email:
h.farahani@umz.ac.ir}\hspace{1mm}}\\
{\small {\em  Department of physics, Mazandaran University,
Babolsar, Iran}}\\
{\small {\em P .O .Box 47416-95447, Babolsar, Iran}} } \maketitle
\begin{abstract}
In this paper we study thermodynamics, statistics and spectroscopic
aspects of a charged black hole with a scalar hair coupled to the
gravity in (2+1) dimensions. We obtained effects of the black hole
charge and scalar field on the thermodynamical and statistical
quantities. We find that scalar charge may increase entropy,
temperature and probability, while may decrease black hole mass,
free and internal energy. Also electric charge increases probability
and decreases temperature and internal energy.
Also we investigate stability of the system and find that the thermodynamical stability exists.
\end{abstract}
\newpage
\tableofcontents
\newpage
\section{Introduction}
Recently charged black hole with a scalar hair in (2 + 1)
dimensions, where the scalar field couples to gravity and it also
couples to itself with the self-interacting potential, analyzed by
the Xu and Zhao [1]. They found appropriate conditions for the
metric in which a charged extremal black hole reproduced as an
asymptotically AdS space-time with a naked singularity at the
origin. Also they shown that the size of the outer horizon increases
monotonically with both the scalar hair and the electric charge.
Refs. [2-6] has similar studies on (2+1)-dimensional rotating hairy
black holes. Such 3D black holes are also interesting to study
AdS/CFT and related topics [7, 8, 9]. However calculation of the
thermodynamic quantities left as an open problem. We know that the
black hole thermodynamics and statistics are interesting fields of
theoretical physics. So we studied in several papers these subjects
for various black holes [10-17]. In that case critical behaviors of
thermodynamical quantities of 3D black holes with a scalar hair has
been studied in the Ref. [18] and concluded the presence of the
cosmological constant may be considered as a thermodynamic pressure
and its conjugate quantity as a volume. Also it is found that the
corresponding equation of state predicts a critical universal number
depending on scalar hair parameter.\\
Now, in this paper we consider charged black hole with a scalar hair
in (2 + 1) dimensions and calculate thermodynamical and statistical
quantities. Indeed we investigate effects of black hole charge and
scalar field on thermodynamical and statistical parameters. Also we
study spectroscopic aspects [19-21] of this black hole and obtain
discrete spectrum. We find that this system has thermodynamical
stability. Therefore in next section we review charged hairy black
hole in (2+1) dimensions. Then in section 3 we study thermodynamics
and obtain entropy, temperature and heat capacity which yields to
obtain thermal stability. spectroscopic aspects of this black hole
studied in section 4 and statistical mechanics of this black hole
investigated in section 5. Finally in section 6 we give conclusion.
\section{Charged hairy black hole in (2+1) dimensions}
In this work we consider a class of black hole solution obtained by
the Ref. [1]. The solution represents static charged black hole with
a scalar field given by the following action,
\begin{equation}\label{s1}
S=\frac{1}{2}\int{d^{3}
x\sqrt{-g}[R-g^{\mu\nu}\nabla_{\mu}\phi\nabla_{\nu}\phi-\xi
R\phi^2-2V(\phi)-\frac{1}{4}F_{\mu\nu}F^{\mu\nu}]},
\end{equation}
where $\xi$ is a coupling constant between gravity and the scalar
field which will be fixed as $\xi=1/8$ [1].\\
This black hole described by the following static, circularly and
asymptotic metric [1],
\begin{eqnarray}\label{s2}
ds^2=-f(r)dt^2+\frac{1}{f(r)}dr^2+r^{2}d\psi^{2},
\end{eqnarray}
where,
\begin{equation}\label{s3}
f(r)=(3\beta-\frac{Q^2}{4})+(2\beta-\frac{Q^2}{9})\frac{B}{r}-Q^2(\frac{1}{2}+\frac{B}{3r})\ln(r)+\frac{r^2}{l^2}.
\end{equation}
$Q$ is the electric charge, $l$ is integration constants and related
to the cosmological constant, $r$ denotes the usual radial
coordinate with $r\geq0$, $-\infty<t<\infty$, and
$-\pi\leq\psi\leq\pi$. Also $\beta$ is integration constants depends
on the black hole charge and mass, and $B$ related to the scalar
field as,
\begin{eqnarray}\label{s4}
\beta&=&\frac{1}{3}(\frac{Q^2}{4}-M),\nonumber\\
\phi(r)&=&\pm\sqrt{\frac{8B}{r+B}}.
\end{eqnarray}
By imposing various constraints, the horizons radius are obtained
[1] and we go ahead the black hole thermodynamics through these
cases.
\section{Thermodynamics}
In order to study thermodynamics of a charged hairy black hole in
(2+1) dimensions we begin with the generic form of the entropy,
\begin{equation}\label{s5}
s=\frac{(g_{\psi\psi})^{\frac{d-1}{2}}}{4G_{N}}|_{r=r_{+}},
\end{equation}
where $r_{+}$ is horizon radius obtained by $f(r)=0$ using the
equation (3). Therefore, using the metric (2) in the relation (5)
gives the following entropy,
\begin{equation}\label{s6}
s=4\pi r_{+},
\end{equation}
where we assumed $16\pi G_{N}=1$. In order to obtain black hole
temperature, potential and special heat capacity we used the
following relations, respectively,
\begin{equation}\label{s7}
T=(\frac{\partial M}{\partial s}),
\end{equation}
\begin{equation}\label{s8}
V=(\frac{\partial M}{\partial Q}),
\end{equation}
and,
\begin{equation}\label{s9}
C=T(\frac{\partial s}{\partial T}),
\end{equation}
where $M$ and $Q$ are the black hole mass and charge respectively.
The sign of specific heat determines thermodynamics stability of
black hole. If the black holes have positive specific heat then it
is in stable equilibrium [22, 23].\\
First of all we study the special case of uncharged black hole.
Then, in order to find the effect of the black hole charge extend
our work to the several cases of charged black hole.
\subsection{Uncharged black hole}
In the case of uncharged black hole we set $Q=0$ in the equation
(3). The other important parameter is $\beta$. In the case of
$\beta=0$ we see that the black hole radius and therefore the black
hole entropy vanishes. On the other hand for the case of
$\beta\neq0$ the equation (3) reduced to the following,
\begin{equation}\label{s10}
f(r)=\frac{r^2}{l^2}-M-\frac{2BM}{3r}.
\end{equation}
Then, the black hole horizon is given by,
\begin{equation}\label{s11}
r_{+}={\frac{\left(9BM{l}^{2}+3\sqrt{3}\sqrt{{M}^{2}{l}^{4}
\left(3B^{2}-M{l}^{2}\right)}\right)^{2/3}+3M{l}^{2}}{3
\left(9BM{l}^{2}+3\sqrt{3}\sqrt{{M}^{2}{l}^{4}\left(3{B}^{2}-M{l}^
{2}\right)}\right)^{\frac{1}{3}}}}.
\end{equation}
Therefore the black hole entropy easily obtained by using the
equation (6). In the Fig. 1 we see that the black hole entropy
increased by scalar field parameter $B$. This is in agreement with
the result of the Ref. [1] which tells that the size of the black
hole must increase as $B$ increases.

\begin{figure}[th]
\begin{center}
\includegraphics[scale=.25]{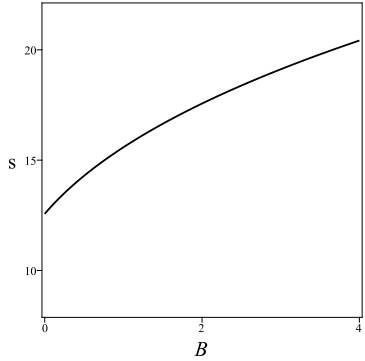}
\caption {Plots of black hole entropy in terms of $B$ with $M=1$ and
$l=1$.}
\end{center}
\end{figure}

Then, using the relations (6) and (10) we can extract
entropy-dependent mass of the black hole as the following,
\begin{equation}\label{s12}
M=\frac{3{s}^{3}}{16{\pi}^{2}{l}^{2}(8\pi B+3s)}.
\end{equation}
By using the relation (7) we can obtain the black hole temperature
in terms of the entropy as the following,
\begin{equation}\label{s13}
T=\frac{9{s}^{2}(4\pi B+s)}{8{\pi}^{2} {l}^{2}(8\pi B+3s)^{2}}.
\end{equation}
In the fig. 2 we can see that the scalar field parameter $B$
decreases the black hole mass but increases the black hole
temperature. Since $\Lambda=-\frac{1}{l^2}$, then we conclude that
increasing of $l$ (or decreasing of cosmological constant) can
reduces the black hole mass and temperature. Finally, from the
equation (9) one can obtain,
\begin{equation}\label{s14}
C=\frac{(8\pi B+3s)s(4\pi B+s)}{3s^{2}+24\pi Bs+64{\pi}^{2}{B}^{2}}.
\end{equation}
In the Fig. 3 we draw specific heat in terms of entropy. We see
thermodynamical stability of this system in any temperature with
varying $B$.

\begin{figure}[th]
\begin{center}
\includegraphics[scale=.25]{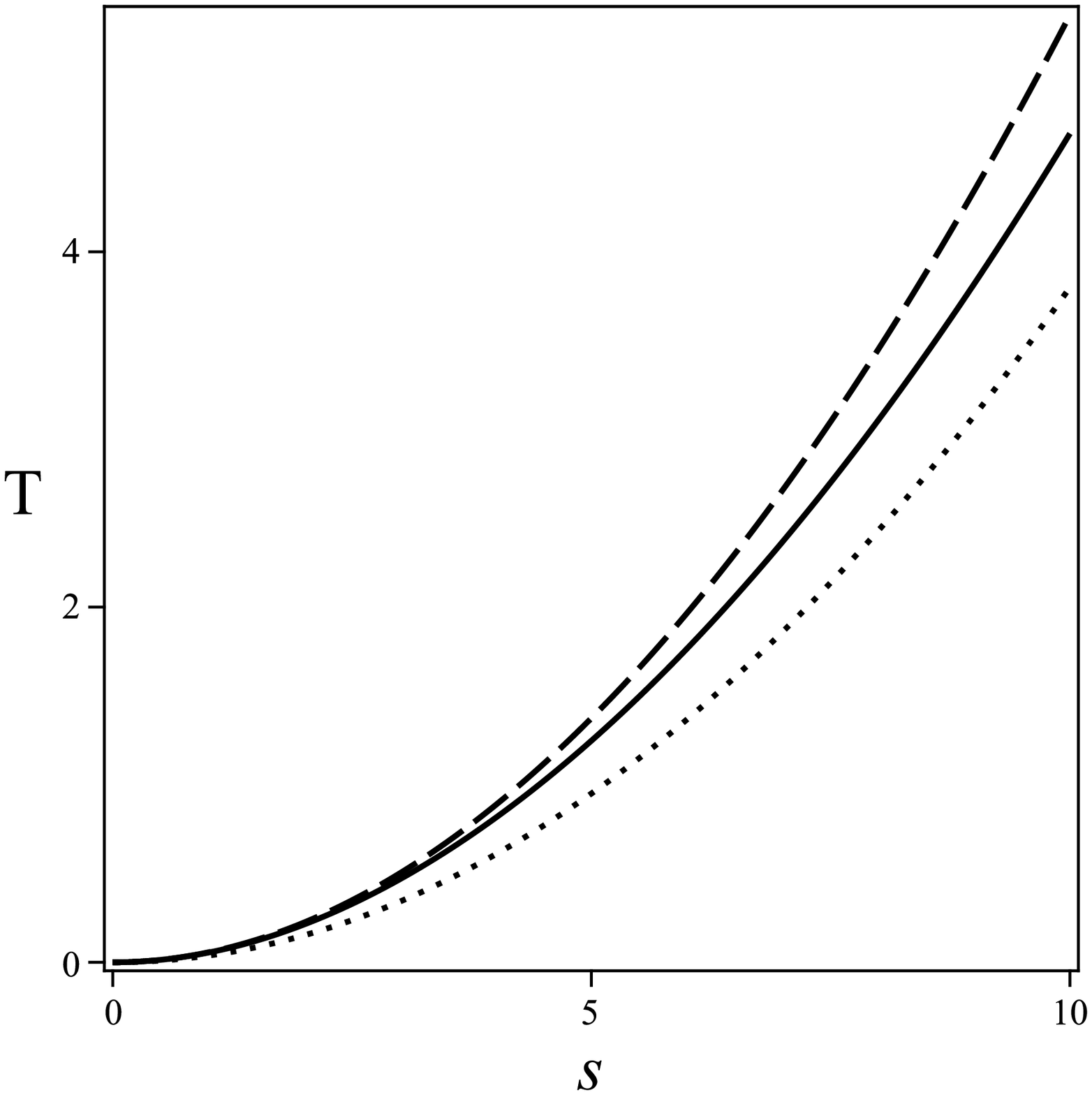}\includegraphics[scale=.25]{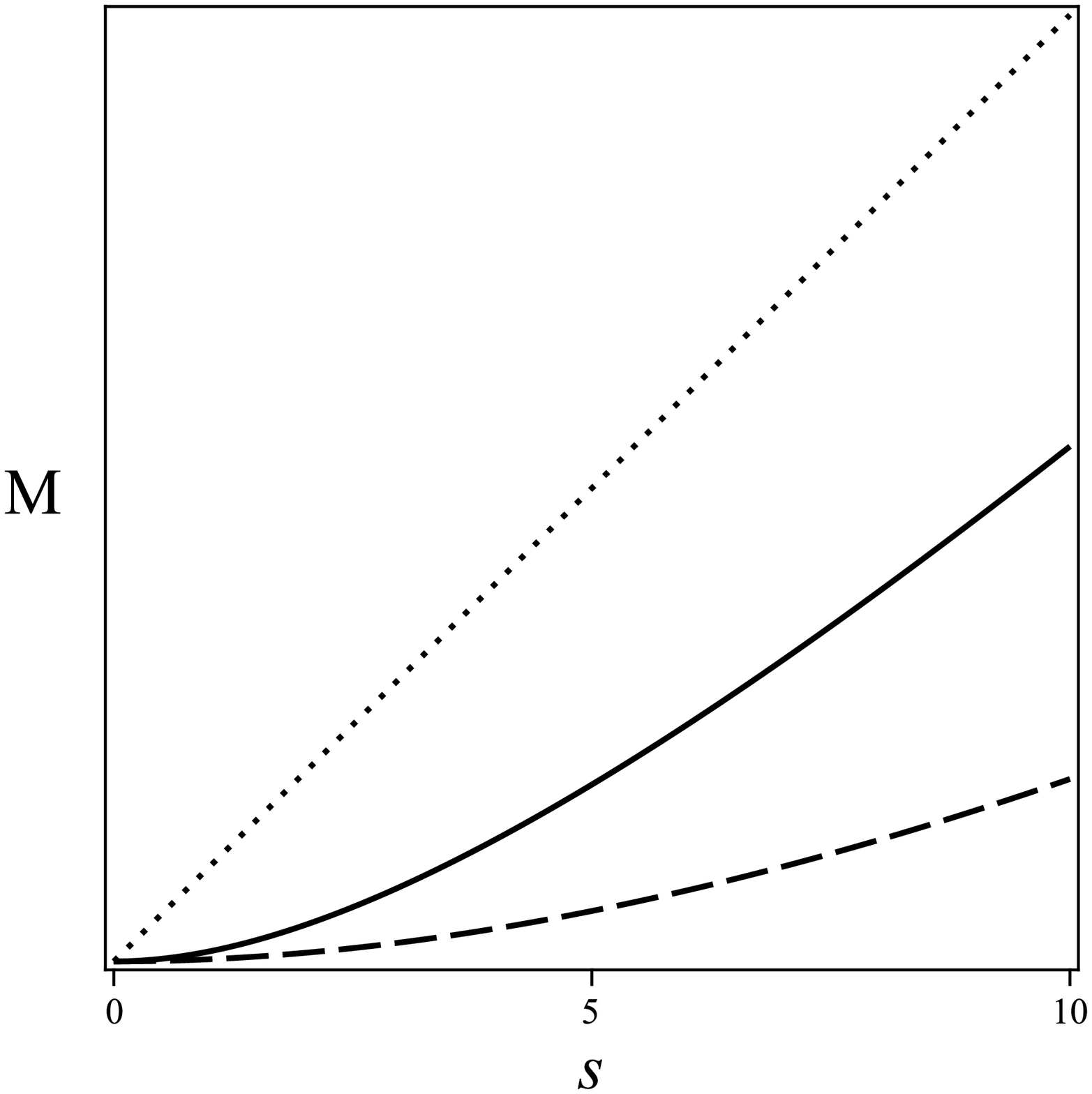}
\caption {Plots of black hole temperature and mass with $l=1$ and
for $B=0$ (dotted line), $B=1$ (solid line), $B=5$ (dashed line).}
\end{center}
\end{figure}

\begin{figure}[th]
\begin{center}
\includegraphics[scale=.25]{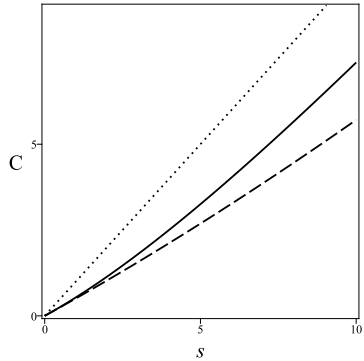}
\caption {Plot of heat capacity versus entropy with $l=1$ and for
$B=0$ (dotted line), $B=1$ (solid line), $B=5$ (dashed line).}
\end{center}
\end{figure}

In the special case of $\beta=-\frac{B^2}{l^2}$, where the self
interacting scalar field vanishes, we have a conformal black hole
with,
\begin{equation}\label{s15}
f(r)={\frac{{r}^{2}}{{l}^{2}}}-{\frac
{3{B}^{2}}{{l}^{2}}}-{\frac{2{B}^{3}}{r{l}^{2}}},
\end{equation}
which yields to the following black hole mass,
\begin{equation}\label{s16}
M=\frac{3s^2}{64\pi^2 l^2}.
\end{equation}
Since $B>0$, the horizon radius is given by
$r_{+}=2B=2l\sqrt{\frac{M}{3}}$, therefore the black hole entropy is
obtained as the following,
\begin{equation}\label{s17}
s=8\pi B =8\pi l\sqrt{\frac{M}{3}}.
\end{equation}
Hence the black hole temperature reads,
\begin{equation}\label{s18}
T={\frac{3}{32}}{\frac{s}{{\pi}^{2}{l}^{2}}},
\end{equation}
where we used the relations (7) and (16). Then the special heat is
given by,
\begin{equation}\label{s19}
C=8\pi l\sqrt{\frac{M}{3}}=\frac{32\pi^2 l^2}{3}T=s.
\end{equation}
It is interesting to find special heat equal to the entropy. It
tells that the heat capacity never decreases and our system will be
in thermal stability.

\subsection{Charged black hole}
Our interesting case is the charged black hole where we can
investigate the effect of black hole charge on thermodynamical
quantities. We will consider two special cases depend on scalar
field parameter $B$ as the following.
\subsubsection{The case of $B=0$}
If we set $B=0$ then the scaler field vanishes and the charged BTZ
black hole appears in (2+1) dimensions reproduced [24, 25]. In that
case the relation (3) reduced to the following,
\begin{equation}\label{s20}
f(r)={\frac {{r}^{2}}{{l}^{2}}}-M-\frac{Q^{2}}{2}\ln(r).
\end{equation}
Now, we discuss about various ranges of $\beta$ which play an
important role on the space-time.\\
In the case of $\beta<\frac{Q^2}{6}\ln(r)$, the $f(r)$ has two zeros
that each of them correspond to the black hole horizons, then there
is a non-extremal charged AdS black hole with no scalar hair.\\
In the case of $\beta=\frac{Q^2}{6}\ln(r)$, the $f(r)=0$ obtained
for the $r=0$.\\
In the case of $\beta>\frac{Q^2}{6}\ln(r)$, the $f(r)$ has no zero
that causes an asymptotically AdS space-time with a naked
singularity in the origin.\\
The extremum of the metric function (20), which is obtained by
$f^{\prime}(r)=0$, given for the radius $r=\frac{1}{2}Ql$. In that
case the black hole mass is given by the following relation,
\begin{equation}\label{s21}
M=\frac{r_{+}^{2}}{l^2}-\frac{Q^2}{2}\ln(r_{+})=\frac{s^2}{16\pi^{2}l^2}-\frac{Q^2}{2}
\ln(\frac{s}{4\pi}).
\end{equation}
Therefore, by using the relation (7) the black hole temperature
obtained as,
\begin{equation}\label{s22}
T=\frac{s}{8\pi^2 l^2}-\frac{Q^2}{2s}.
\end{equation}
Finally the specific heat reads,
\begin{equation}\label{s23}
C=4\pi^2 l T(1+l\sqrt{2T}).
\end{equation}

\begin{figure}[th]
\begin{center}
\includegraphics[scale=.25]{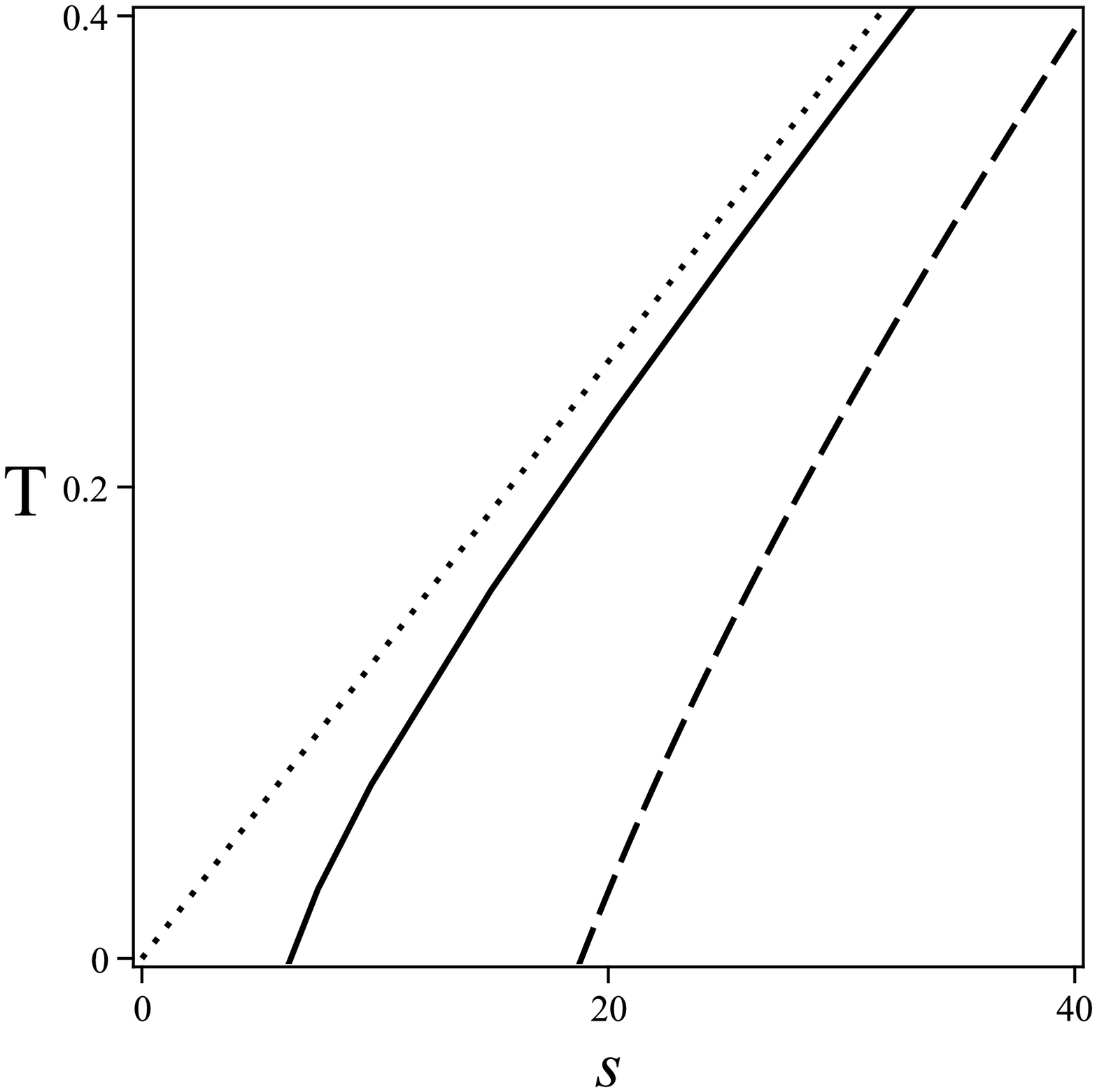}\includegraphics[scale=.25]{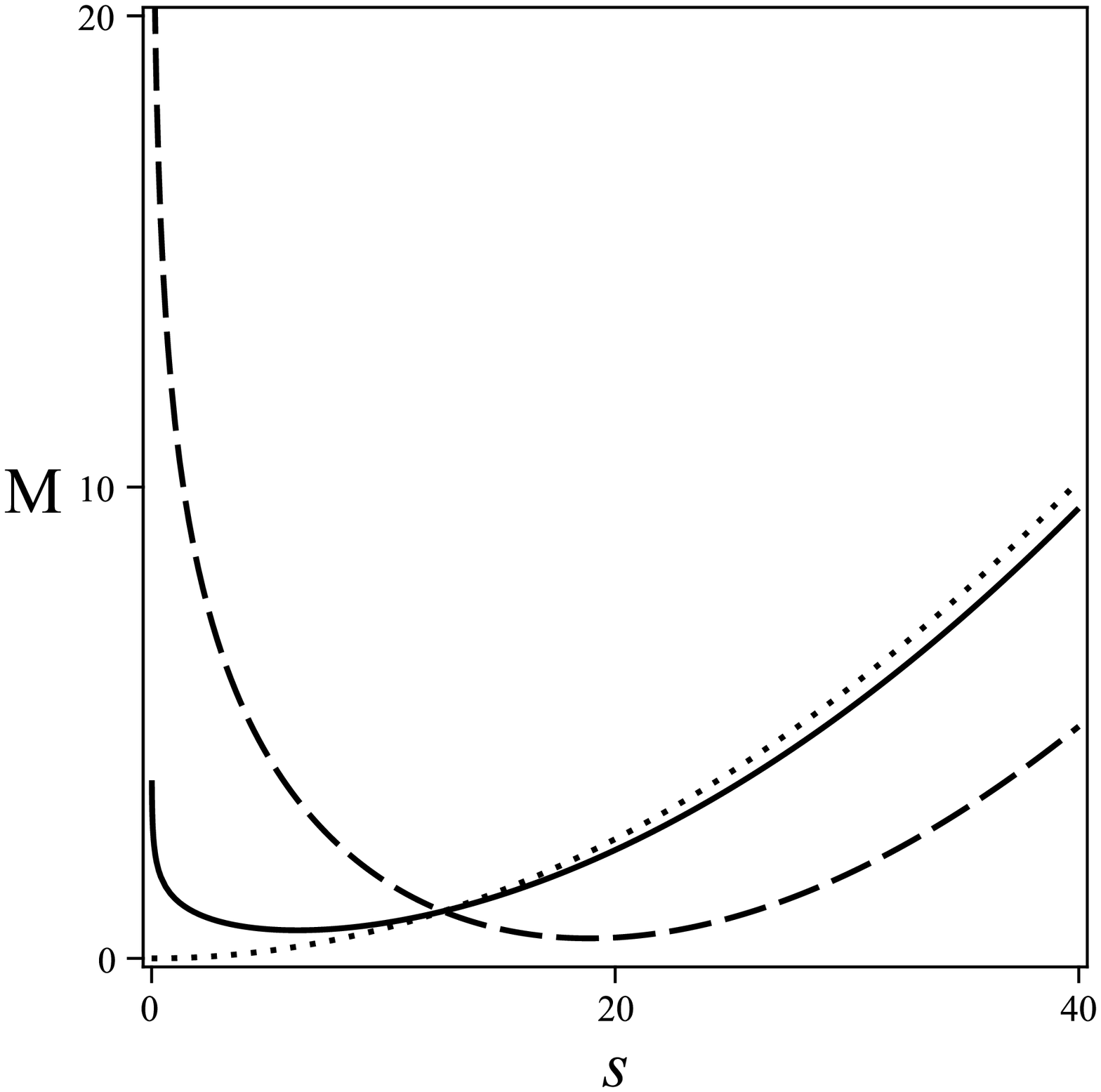}
\caption {Plots of black hole temperature and mass for different
values of black hole charge with $l=1$. $Q=0$ (dotted line), $Q=1$
(solid line), $Q=3$ (dashed line).}
\end{center}
\end{figure}

We find that the black hole charge decreases the black hole
temperature and therefore heat capacity (see Fig. 4). We see that
for higher entropy the black hole charges has no important effect on
the black hole temperature. Also there is a critical point for the
black hole mass ($M_{c}$) where black hole charge increases the
black hole mass for $M<M_{c}$ and decreases the black hole mass for
$M>M_{c}$. The critical value of the black hole mass obtained for
$s\simeq12$ for the given black hole charge.

\subsubsection{The case of $B>0$}
In presence of the scalar field and using definition (4) of $\beta$
the metric function (3) reduced to the following,
\begin{equation}\label{s24}
f(r)=\frac{r^{2}}{l^{2}}-M+\frac{(\frac{Q^{2}}{6}-2M)B}{3r}-\frac{Q^{2}}{2}(1+{\frac{2B}{3r}})
\ln(r).
\end{equation}
Then, the black hole mass is given by,
\begin{equation}\label{s25}
M=\frac{9s^{3}+32{Q}^{2}{l}^{2}\pi^{3}B-72{Q}^{2}\pi^{2}{l}^{2}s\ln(\frac{s}{4\pi})
-192{Q}^{2}\pi^{3}{l}^{2}B\ln(\frac{s}{4\pi})}
{48\pi^{2}l^{2}(3s+8\pi B)}.
\end{equation}
Hence, we can obtain the black hole temperature as the following,
\begin{equation}\label{s26}
T=\frac{9s^{4}+36\pi
B{s}^{3}-36\pi^{2}l^{2}Q^{2}s^{2}-208\pi^{3}l^{2}Q^{2}Bs-256\pi^{4}l^{2}Q^{2}B^{2}}{8\pi^{2}l^{2}s(3s+8\pi
B)^{2}}.
\end{equation}
We can see from the Fig. 5 that the scalar charge decreases the
black hole temperature and mass. Also if $B>3$ the black hole mass
has negative sectors, therefore we can find a constraint on the
scalar charge.

\begin{figure}[th]
\begin{center}
\includegraphics[scale=.25]{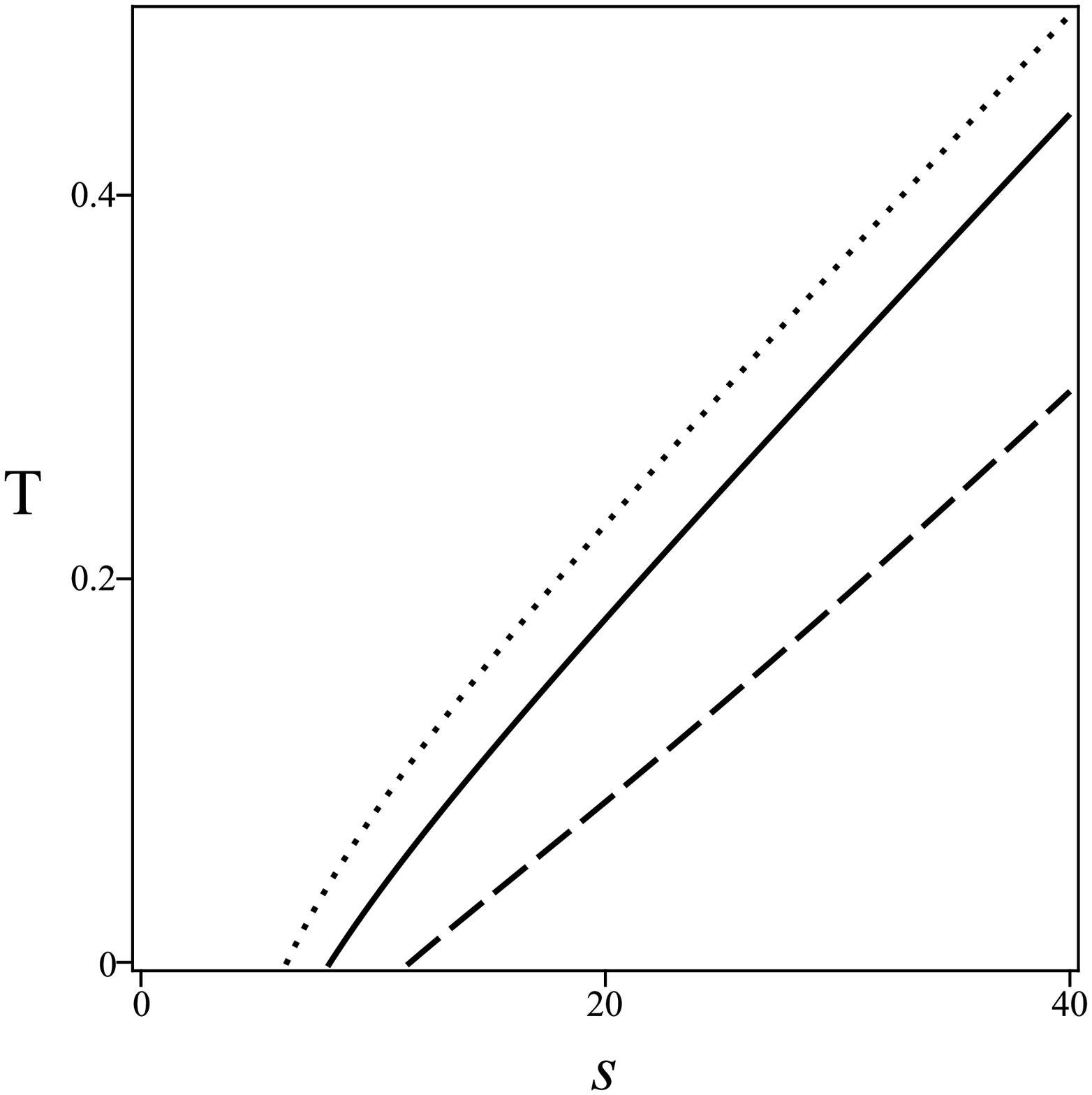}\includegraphics[scale=.25]{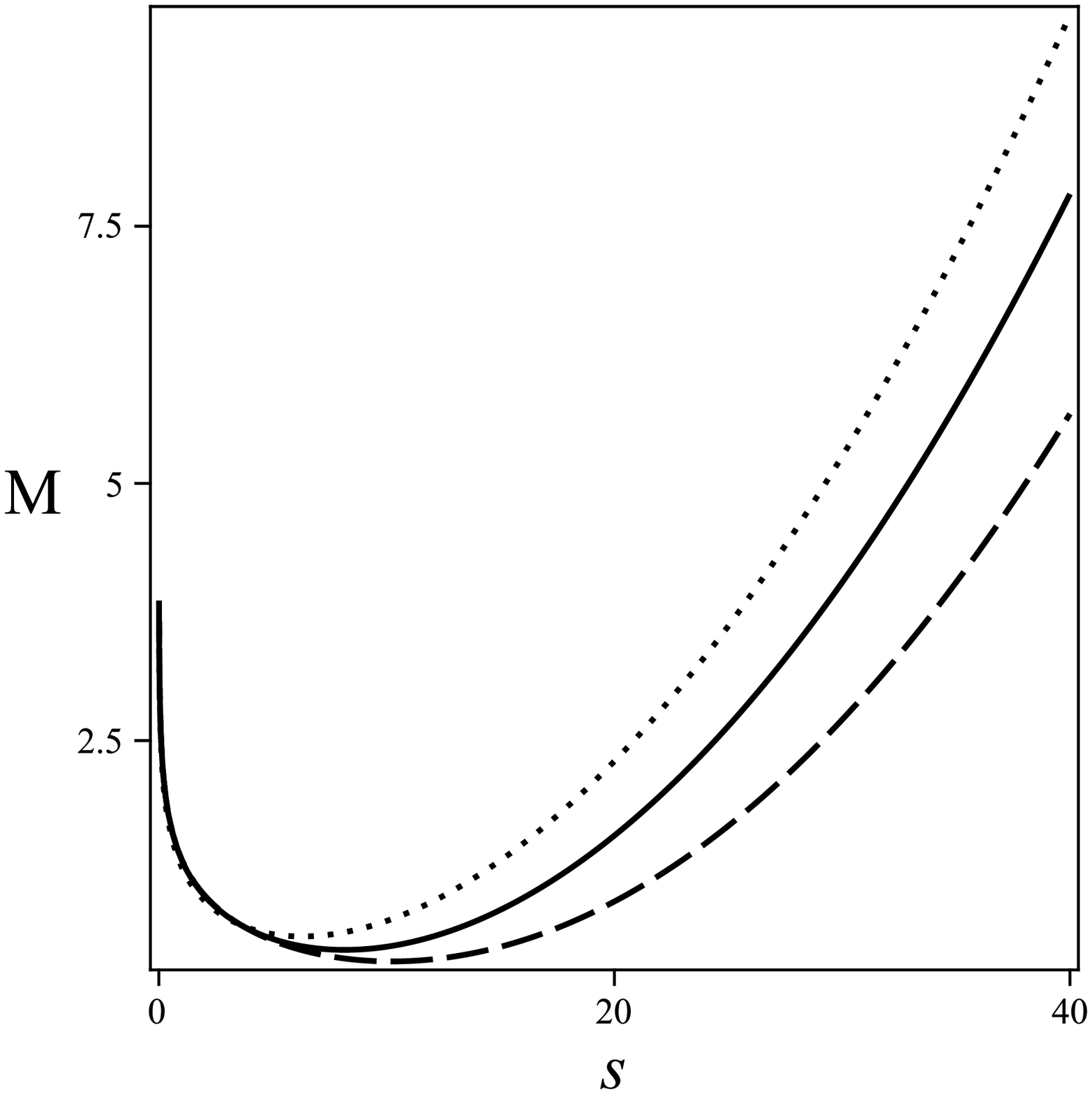}
\caption {Plots of black hole temperature and mass for different
values of the scalar charge with $l=1$ and $Q=1$. $B=0$ (dotted
line), $B=1$ (solid line), $B=3$ (dashed line).}
\end{center}
\end{figure}

Finally the specific heat obtained by using the relation (9). We
find that the scalar charge decreases the heat capacity and there
are some negative sectors, however for appropriate choices of $B$,
$Q$ and $s$ where the black hole temperature is positive then, the
system is in thermal stability (see Fig. 6).

\begin{figure}[th]
\begin{center}
\includegraphics[scale=.25]{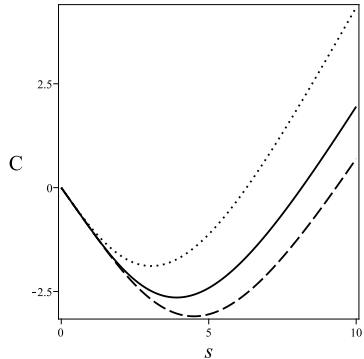}
\caption {Plots of heat capacity for different values of the scalar
charge with $l=1$ and $Q=1$. $B=0$ (dotted line), $B=1$ (solid
line), $B=2$ (dashed line).}
\end{center}
\end{figure}

\section{Spectroscopy}
In this section we discuss the properties of entropy quantization
and the area spectrum. In that case the first law of thermodynamics
for charged black hole is [26],
\begin{equation}\label{s27}
dM=\frac{T}{4}dA+VdQ,
\end{equation}
where $A$ is surface area of the event horizon which for the (2+1)
dimensional black hole is given by,
\begin{equation}\label{s28}
A=2\pi r_{+}.
\end{equation}
Therefore we can write,
\begin{equation}\label{s29}
\Delta A=2\pi dr_{+}.
\end{equation}
We try to obtain circumference spectrum for two different cases of
uncharged and charged black holes.

\subsection{Uncharged black hole}
In the first case we set $Q=0$. Then, for the case of $\beta\neq0$,
the equations (13), (27) and (29) give the following expression,
\begin{equation}\label{s30}
dM=\frac{9}{l^{2}}\frac{r_{+}^{2}}{2B+3r_{+}}\left[1-\frac{r_{+}}{2B+3r_{+}}\right]dr_{+}.
\end{equation}
Using the lapse function we can calculate the surface gravity as the
following,
\begin{equation}\label{s31}
\kappa=\frac{1}{2}\frac{df(r)}{dr}|_{r=r_{+}}=\frac{8r_{+}-1}{8l^{2}}.
\end{equation}
Then, using the equation (30) we can obtain,
\begin{equation}\label{s32}
dr_{+}=\frac{4}{9}\frac{l^{2}(16B+3+24\kappa l^{2})^{2}}{(1+8\kappa
l^{2})^{2}(8B+1+8\kappa l^{2})}dM.
\end{equation}
Therefore we one can find,
\begin{equation}\label{s33}
\Delta A=\frac{8\pi}{9}\frac{l^{2}(16B+3+24\kappa
l^{2})^{2}}{(1+8\kappa l^{2})^{2}(8B+1+8\kappa l^{2})}dM,
\end{equation}
where we used the equations (29) and (32). Now, by replacing
$\kappa$ in terms of $M$, the adiabatic invariant integral is given
by the following,
\begin{equation}\label{s34}
I=\int \Delta A=\frac{32\sqrt{3}\pi}{27}\sqrt{M},
\end{equation}
which is achieved by supposing the special case of
$B=\sqrt{\frac{M}{3}}$ and
$l=1$.\\
Applying Bohr-Sommerfeld quantization condition and the equation in
terms of $r_{+}$ one can obtain,
\begin{equation}\label{s35}
I=\frac{8}{9} (2\pi r_{+})\simeq n\hbar.
\end{equation}
Therefore we can obtain,
\begin{equation}\label{s36}
A_{n}\simeq \frac{9}{8}n\hbar.
\end{equation}
We can we see that the spectrum of uncharged hairy black hole is
discrete, and the spectrum is independent of the black hole
parameters as expected.

\subsection{Charged black hole}
First of all we study the case of $B=0$ and obtain,
\begin{equation}\label{s37}
\Delta A=4\left(\frac{r_{+}}{2\pi l^{2}}-\frac{Q^{2}}{8\pi
r}\right)^{-1}(dM+\frac{dQ}{l^{2}}),
\end{equation}
where we used $V=-1/l^{2}$ [1], also horizon radius in terms of
black hole mass and charge obtained as the following,
\begin{equation}\label{s38}
r_{+}=\exp\left(-\frac{1}{2}LambertW\left[-\frac{4e^{-\frac{4M}{Q^{2}}}}{l^{2}Q^{2}}\right]-\frac{2M}{Q^{2}}\right),
\end{equation}
which reduced to the following relation in asymptotic expansion for
the large $Q$,
\begin{equation}\label{s39}
r_{+}=e^{\frac{2(1-M)}{Q^{2}}},
\end{equation}
where $l=1$ assumed. After some calculations one can obtain,
\begin{equation}\label{s40}
A_{n}\simeq \frac{n\hbar}{8}-\frac{32\pi}{Q}
\end{equation}
As before for the first term we see that the circumference of the
charged hairy black hole is discrete. In this system, the
circumference spectrum depends on the black hole parameters, but the
circumference spacing is independent of the black hole parameters.\\
On the other hand by choosing $B=\frac{Ql}{6}$ we can obtain the
surface gravity as the following,
\begin{equation}\label{s41}
\kappa=\frac
{648r_{+}^{4}+108Qlr_{+}^{3}+162{Q}^{2}{l}^{2}r_{+}^{2}-39{Q}^{3}{l}^{3}r_{+}-2{Q}^{4}{l}^{4}}{72{l}^{2}
\left(9r_{+}+ Ql\right) r_{+}^{2}},
\end{equation}
which yields to the following result,
\begin{eqnarray}\label{s42}
A_{n}\simeq 2n\hbar+\frac {\pi
lQ}{5832}\left((68+368\ln(l))Q^{3}+(108+864\ln(l)){Q}^{2}+5562\ln(l)
{Q}+1296\right).
\end{eqnarray}
This case of a charged black hole with scalar hair has an
independent spacing circumference and its spectrum is achieved in
respect of $l$ ($B$) and $Q$ as black hole parameters.

\section{Statistical mechanics}
By using the thermodynamical quantities one can obtain some
important quantities such as partition function which is important
in statistical mechanics. First of all we should obtain free energy,
\begin{eqnarray}\label{s43}
F=- \int s dT.
\end{eqnarray}
This quantity is important to study statistics of system and also
useful to obtain internal energy,
\begin{eqnarray}\label{s44}
E=F+ sT.
\end{eqnarray}
Then one can obtain partition function as the following,
\begin{eqnarray}\label{s45}
Z=\exp(-\frac{F}{k T}),
\end{eqnarray}
where $k$ is Boltzmann constant which we assumed $k=1$ for
simplicity. By using partition function one can obtain probability,
\begin{eqnarray}\label{s46}
p=\exp(-\frac{\Xi}{Z}),
\end{eqnarray}
where $\Xi$ is called grand partition function and defined by,
\begin{eqnarray}\label{s47}
\Xi=\exp(-\frac{E}{T}),
\end{eqnarray}
and microcanonical states number is given by,
\begin{eqnarray}\label{s48}
\Omega=-T \ln(\Xi)
\end{eqnarray}
The number of micro states produces statistical entropy as the
following,
\begin{eqnarray}\label{s49}
s_{micro}=ln \Omega.
\end{eqnarray}

\subsection{The case of $Q=0$}
In that case we have two separated cases depend on parameter
$\beta$. In the first case we set $\beta=0$, where the entropy
obtained by using the relation (11). By using the relation (43) and
(44) we draw free energy and internal energy in the Fig. 7. It shows
that the free energy reduced by scalar charge and internal energy
has a maximum at $B=0$.

\begin{figure}[th]
\begin{center}
\includegraphics[scale=.25]{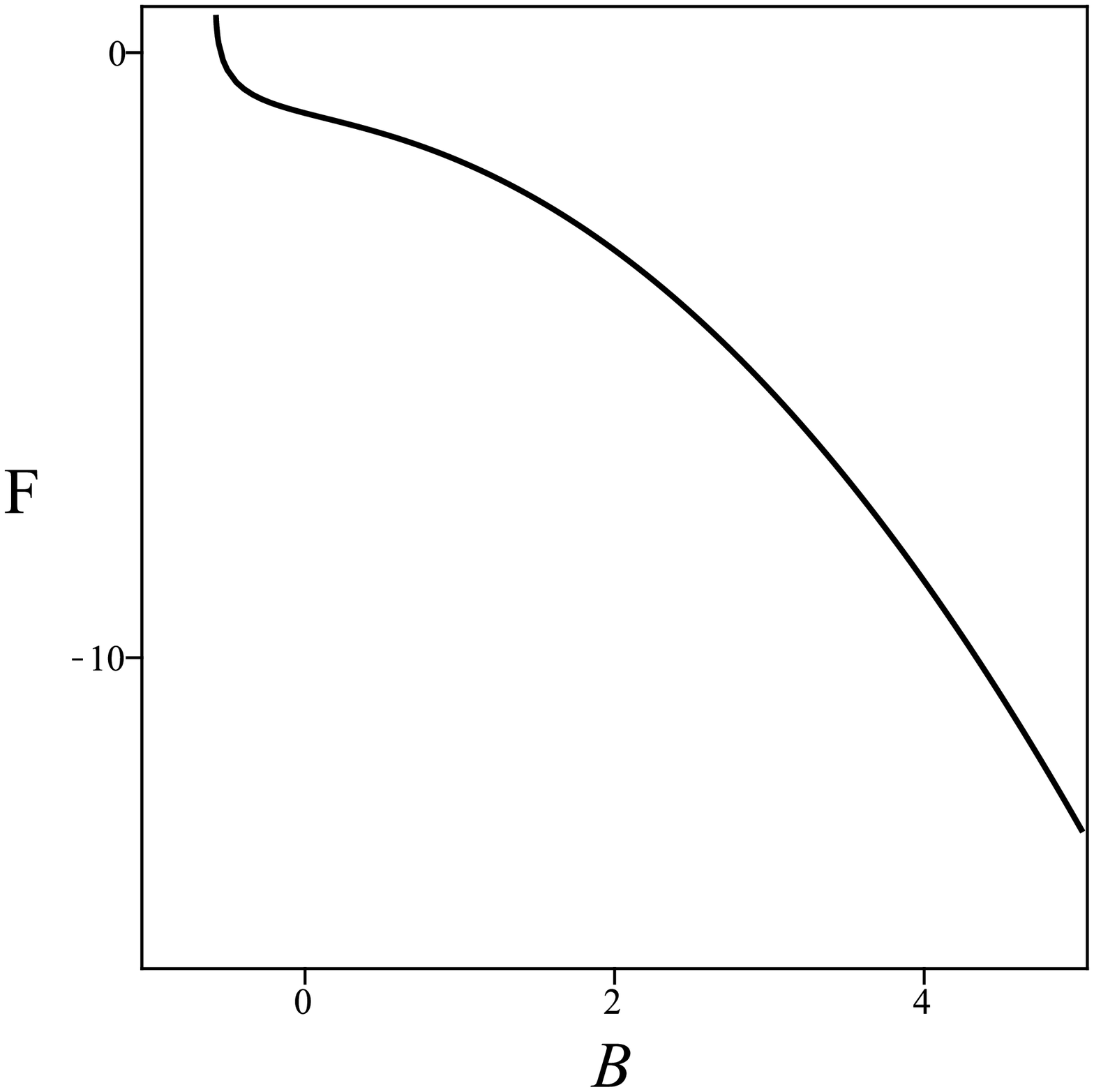}\includegraphics[scale=.25]{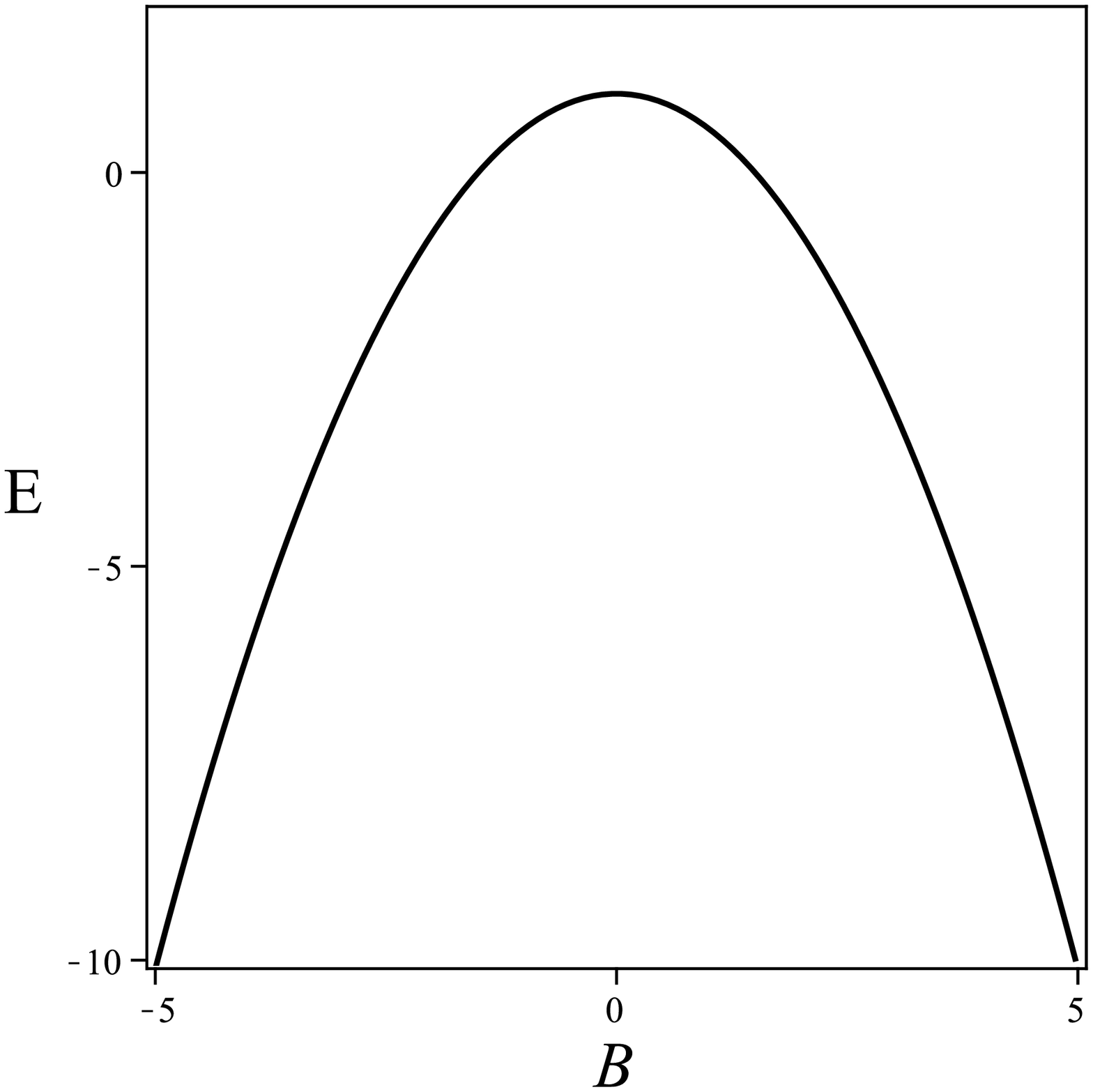}
\caption {plots of free energy and internal energy in terms of
scalar charge with $M=1$ and $l=1$.}
\end{center}
\end{figure}

Therefore we can obtain partition function as the following,
\begin{equation}\label{s50}
Z=e^{\frac{2\pi\left(81r_{+}^{4}+108B r_{+}^{3}+36
B^{2}r_{+}^{2}+48B^{3}r_{+}+16B^{4} \right)}{81\left(B+r_{+}
 \right)r_{+}^{2}}}.
\end{equation}
Also, probability obtained by using the relation (46) as the
following,
\begin{equation}\label{s51}
p=e^{-{e^{-4\pi r_{+}}}},
\end{equation}
which illustrated in the Fig. 8. It shows that maximum probability
available for higher scalar charge.

\begin{figure}[th]
\begin{center}
\includegraphics[scale=.25]{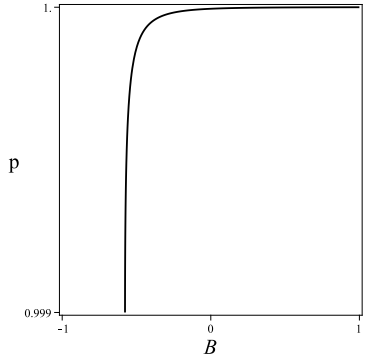}
\caption {plot of probability in terms of scalar charge with $M=1$
and $l=1$.}
\end{center}
\end{figure}

In the second case we set $\beta=-\frac{B^2}{l^2}$ and find that,
\begin{equation}\label{s52}
Z=e^{4\pi B},
\end{equation}
and
\begin{equation}\label{s53}
\Omega=3{\frac {B^{2}}{{l}^{2}}}.
\end{equation}
Therefore we can obtain microcanonical entropy as the following,
\begin{equation}\label{s54}
s_{micro}=\ln(\frac{3B^{2}}{l^{2}}).
\end{equation}
We can see that the microcanonical entropy increased by $B$ as
expected.
\subsection{The case of $Q\neq0$}
In order to obtain the effect of black hole charge we consider two
special cases of $B=0$ and $B\neq0$. In the first case which horizon
radius is given by the equation (38), then the free energy obtained
as,
\begin{equation}\label{s55}
F=r_{+}^{2}+\frac{1}{2}Q^{2}\ln(r_{+}),
\end{equation}
It is easy to check that the black hole charge increased the free
energy. In the Fig. 9 we draw plot of internal energy and find that
there is a maximum for $Q\sim2$.

\begin{figure}[th]
\begin{center}
\includegraphics[scale=.25]{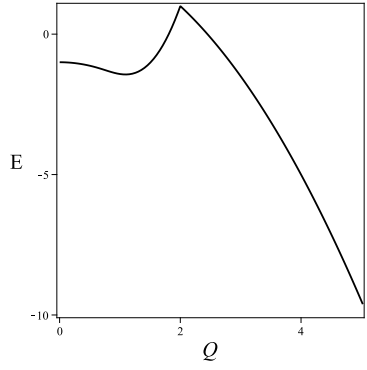}
\caption {plot of internal energy in terms of electric charge with
$M=1$ and $l=1$.}
\end{center}
\end{figure}

Also partition function obtained as the following,
\begin{equation}\label{s56}
Z=1+\frac{8\pi(2-M)}{Q^{2}}+\mathcal{O}(\frac{1}{Q^{4}}),
\end{equation}
where we used asymptotic expansion for the large black hole charge.
Probability of this case illustrated in the Fig. 10.

\begin{figure}[th]
\begin{center}
\includegraphics[scale=.25]{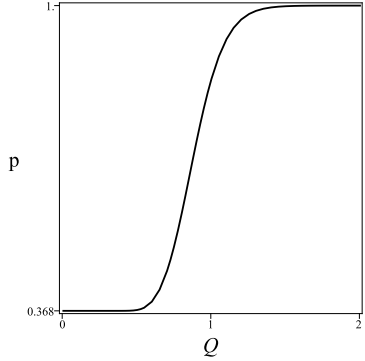}
\caption {plot of probability in terms of black hole charge with
$M=1$ and $l=1$.}
\end{center}
\end{figure}

In the general case we consider $Q\neq0$ and $B\neq0$ and obtain,
\begin{eqnarray}\label{s57}
F=&-&r_{+}^{2}+\frac{1}{3}\frac{BQ}{3r_{+}+2B}(1-\frac{Q}{3r_{+}+2B})\nonumber\\
&-&\frac{16}{9}\frac{B^{3}}{3r_{+}+2B}(1-\frac{B}{3r_{+}+2B})-\frac{1}{2}Q^{2}\ln(r_{+}).
\end{eqnarray}
In order to find effect of electric and scalar charges we draw plot
of free energy in the Fig. 11. In the Fig. 11 (a) we fixed scalar
charge and vary electric charge and find that the electric charge
increases free energy for $r_{+}<1$ and decreases free energy for
$r_{+}>1$. Then in the Fig. 11 (b) we fixed electric charge and vary
scalar charge and find that free energy decreased by scalar charge
as expected from previous results. Similar results obtained for
internal energy $E$.

\begin{figure}[th]
\begin{center}
\includegraphics[scale=.25]{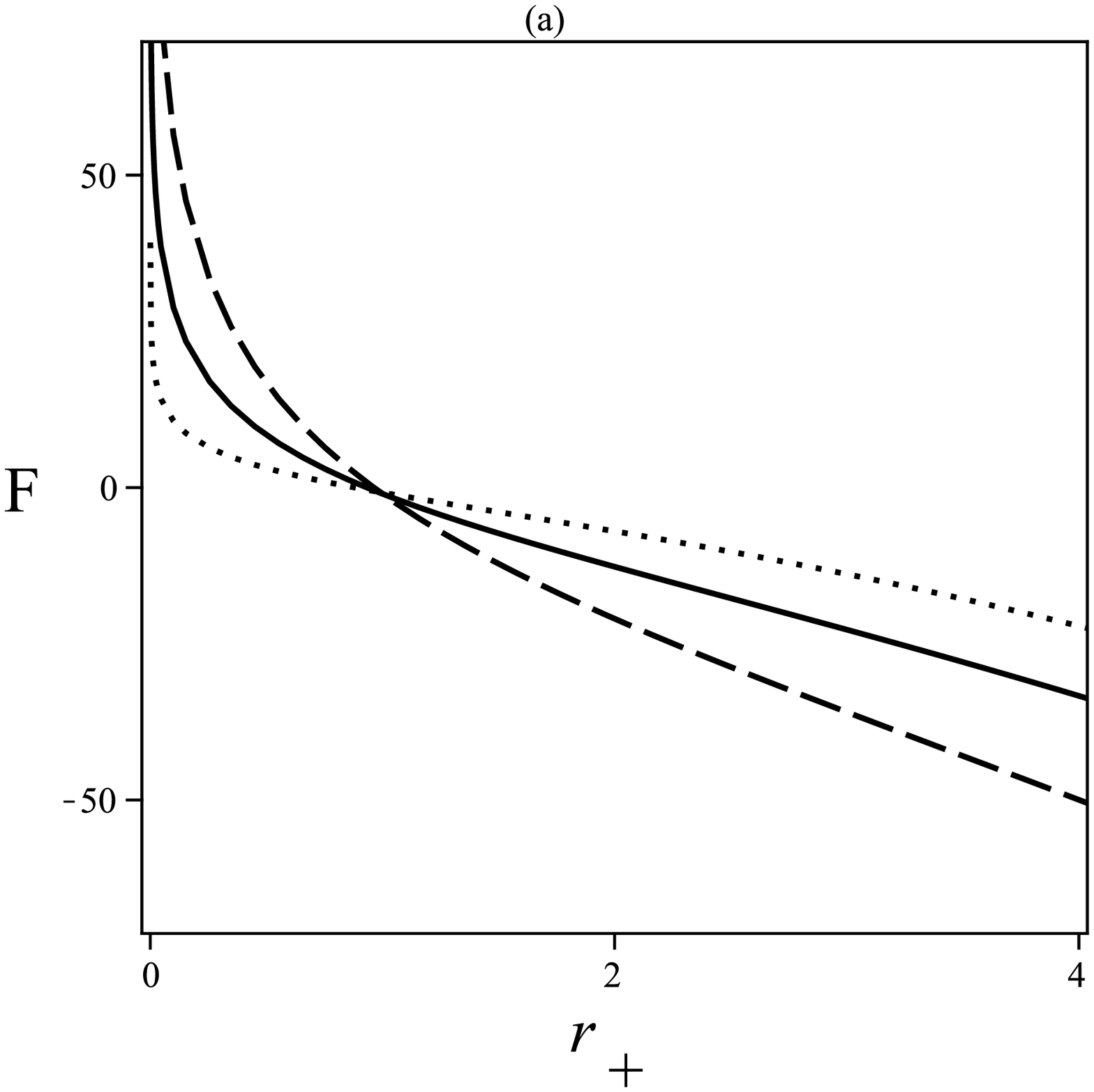}\includegraphics[scale=.25]{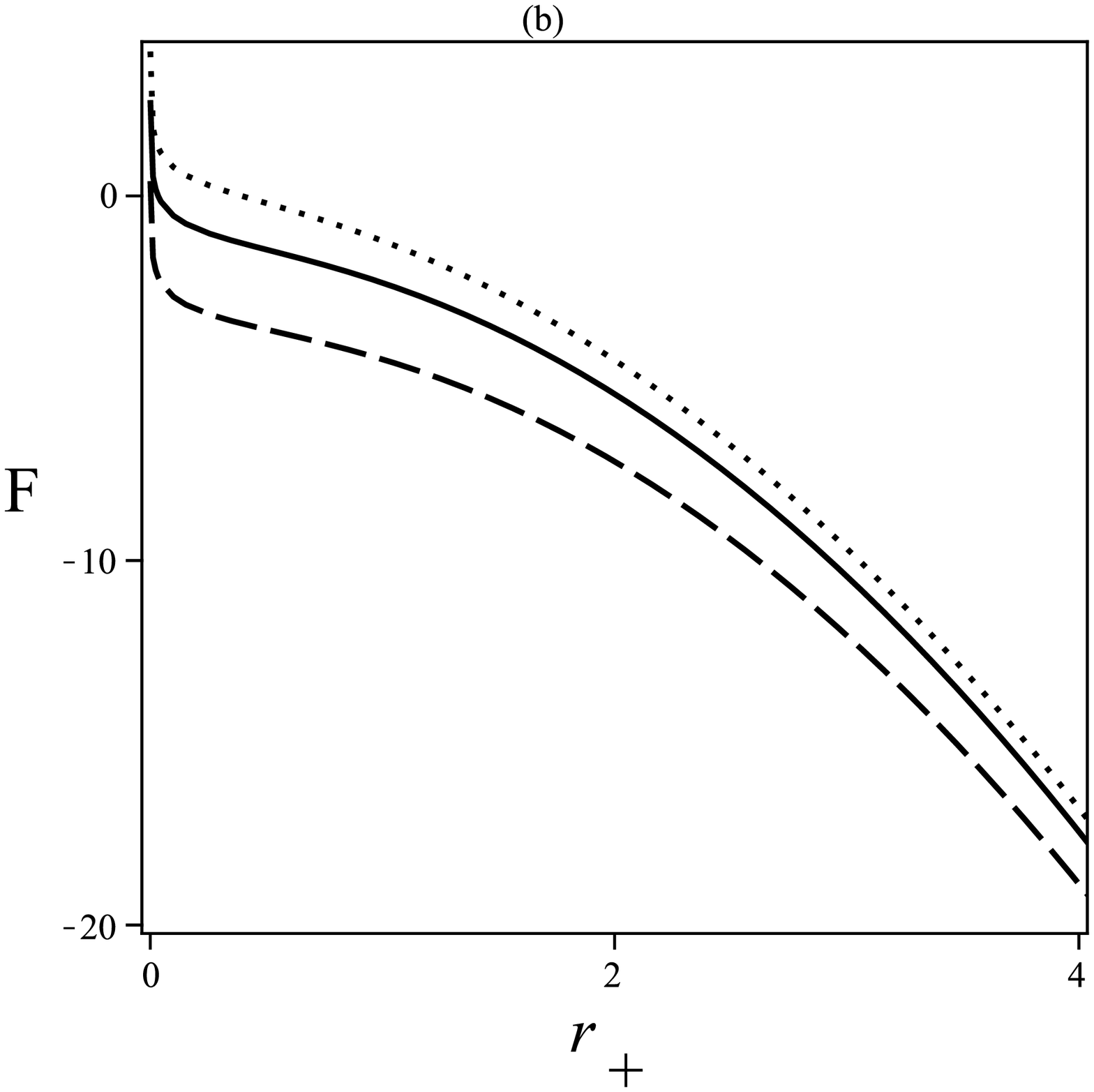}
\caption {plot of free energy in terms of black hole horizon. (a)
$B=1$ and $Q=3$ (dotted line), $Q=5$ (solid line), $Q=7$ (dashed
line). (b) $Q=1$ and $B=1$ (dotted line), $B=2$ (solid line), $B=3$
(dashed line).}
\end{center}
\end{figure}

Probability of this case is also obtained as before. Finally we can
obtain microcanonical entropy which is illustrated in the Fig. 12.
It shows that the electric and scalar charges decrease the
microcanonical entropy.

\begin{figure}[th]
\begin{center}
\includegraphics[scale=.25]{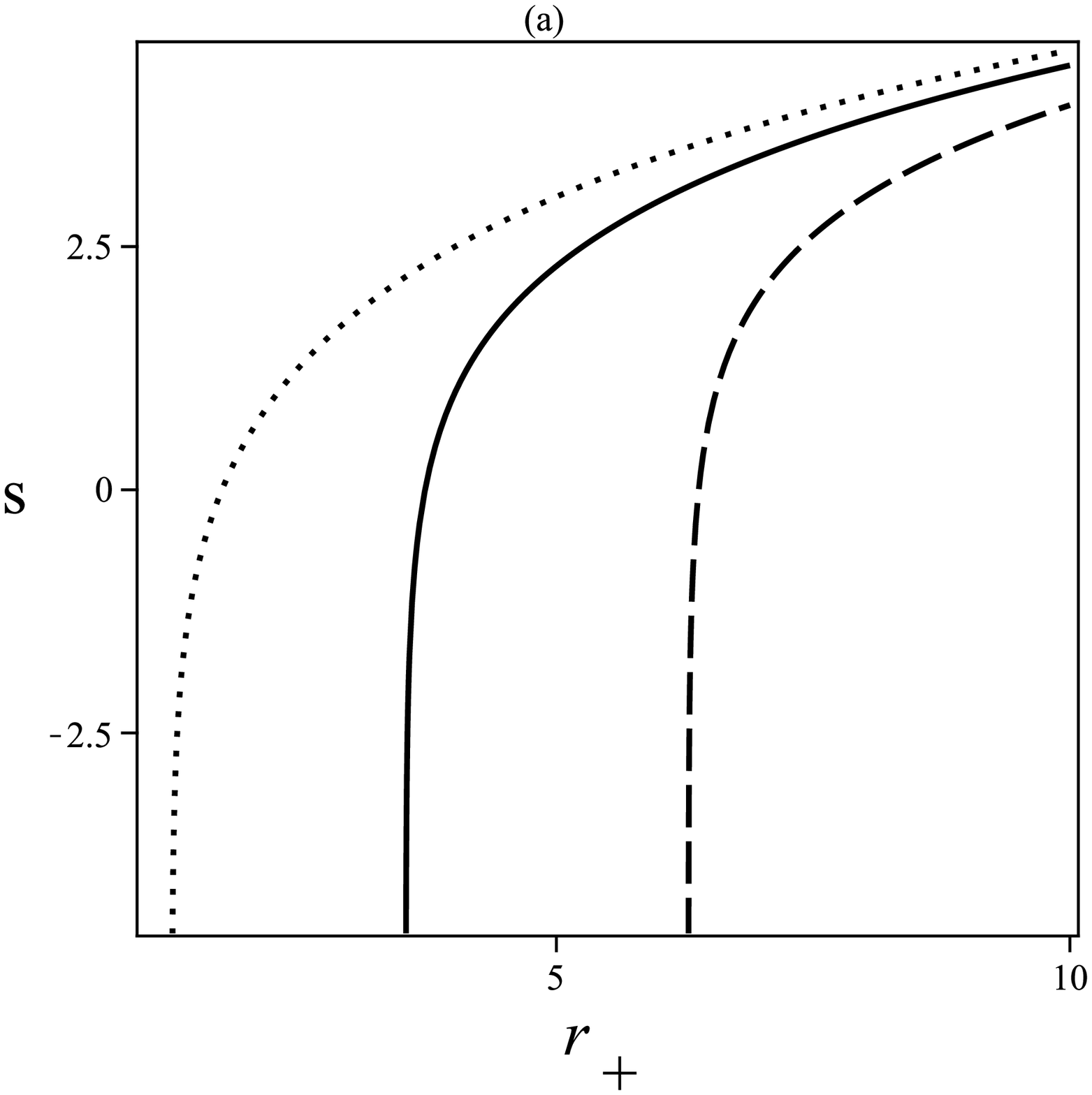}\includegraphics[scale=.25]{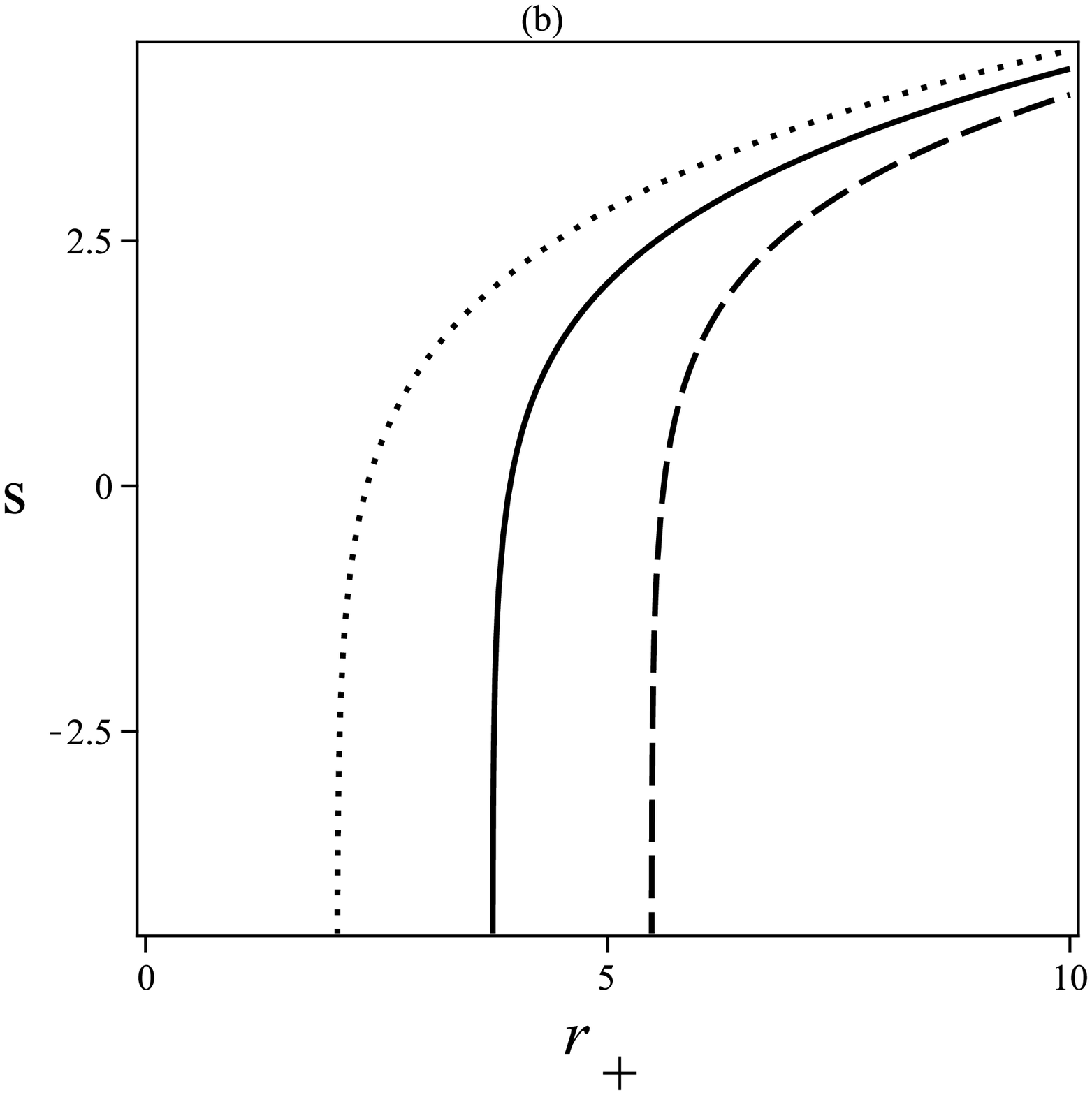}
\caption {plot of microcanonical entropy in terms of black hole
horizon. (a) $B=1$ and $Q=1$ (dotted line), $Q=3$ (solid line),
$Q=5$ (dashed line). (b) $Q=1$ and $B=2$ (dotted line), $B=4$ (solid
line), $B=6$ (dashed line).}
\end{center}
\end{figure}

\section{Conclusions}
In this paper we considered charged hairy black hole in (2+1)
dimensions and studied thermodynamics, statistics and spectroscopy.
In the simplest case of zero-charge black hole we obtained black
hole entropy and found that scalar charge increased it. We obtained
black hole mass and temperature which are increasing function of
entropy. We found that scalar charge increased the black hole
temperature and decreased black hole mass. Then we obtained the
effect of black hole charge on thermodynamics quantities. Similar to
the uncharged case, the black hole temperature is increasing
function of entropy, but black hole mass decreased first and then
increased. In all cases we have thermal stability which verified by
using positive sign of specific heat. Spectroscopy of this system
shows that the circumference of the charged hairy black hole is
discrete. In this system, the circumference spectrum depends on the
black hole parameters, but the circumference spacing is independent
of the black hole parameters. We obtained also partition function
and studied statistical mechanics of charged hairy black hole in
(2+1) dimensions. We found that the microcanonical entropy is
decreasing function of black hole charge.

\end{document}